\documentclass[submission,copyright,creativecommons]{eptcs}
\providecommand{\event}{TFPIE 2017} 
\usepackage{breakurl}             
\usepackage{underscore}           
\usepackage{graphicx}
\usepackage{verbatim}
\usepackage{hyperref}
\usepackage{amsmath}
\usepackage{array}
\usepackage{booktabs,adjustbox}

\title{Using Elm to Introduce Algebraic Thinking to K-8 Students}
\author{Curtis d'Alves, Tanya Bouman, Christopher W. Schankula, Jenell Hogg,\\ Levin Noronha, Emily Horsman, Rumsha Siddiqui, Christopher Kumar Anand
\institute{McMaster University\\ Hamilton, Ontario, Canada}
\email{\{dalvescb,boumante,schankuc,anandc\}@mcmaster.ca}
}
\def\titlerunning{Elm and Algebraic Thinking}
\def\authorrunning{d'Alves, Bouman, Schankula, Hogg, Noronha, Horsman, Siddiqui \& Anand}
\def\copyrightholders{d'Alves, Bouman, Schankula, Hogg,\\ Noronha, Horsman, Siddiqui \& Anand}
\begin{document}
\maketitle

\begin{abstract}
In recent years, there has been increasing interest in developing a Computer Science curriculum for K-8 students.
However, there have been significant barriers to creating and deploying a Computer Science curriculum in many areas, including teacher time and the prioritization of other 21st-century skills.
At McMaster University, we have developed both general computer literacy activities
and specific programming activities.
Integration of these activities is made easy as they each support existing curricular goals.
In this paper, we focus on programming in the functional language Elm and the graphics library GraphicSVG.
Elm is in the ML (Meta Language) family, with a lean syntax and easy inclusion of Domain Specific Languages.
This allows children to start experimenting with GraphicSVG as a language for describing shape,
and pick up the core Elm language as they grow in sophistication.
Teachers see children making connections between computer graphics and mathematics within the first hour.
Graphics are defined declaratively, and support aggregation and transformation, 
i.e., Algebra.
Variables are not needed initially, but are introduced as a time-saving feature, which is immediately accepted.  Since variables are declarative, they match students' expectations.
Advanced students are also exposed to State by making programs that react to user taps or clicks. The syntax required to do so
closely follows the theoretical concepts, making it easy for them to grasp.
For each of these concepts, we explain how they fit into the presentations we make to students, like the 5200 children taught in 2016.

Finally, we describe ongoing work on a touch-based Elm editor for iPad, 
which features (1) type highlighting (as opposed to syntax highlighting),
(2) preservation of correct syntax and typing across transformations,
(3) context information (e.g. displaying parameter names for GraphicSVG functions), and
(4) immediate feedback (e.g. restarting animations after every program change).
\end{abstract}

\section{Introduction}

There is a lot of current interest in developing a Computer Science curriculum for K-8 \cite{Schofield:2014:MCM:2538862.2538901}, with curricula being defined by some education authorities, e.g.{} year-by-year goals in the UK \cite{UKCSCurriculum2013}.
At McMaster University, we have developed a number of outreach activities in collaboration with local teachers and their students.
The best activities are ones which students view as play but which teachers recognize as motivating or reinforcing curricular goals.  
This paper introduces a learning environment 
in which children can develop programming skills initially through exploration and then through discovery, 
motivated by the desire to produce interesting vector graphics including animation,
and for advanced students, interaction.
The dependence on Cartesian coordinates and basic geometry is impossible to miss,
but we think the most important impact on the curriculum is how we can naturally introduce variables and functions with the same semantics as they have in algebra.

While we agree that Computer Science should be integrated into the K-8 curriculum,
it is important to remember that most children will not go on to be software developers
but, more so than ever before, every student needs to succeed in secondary education,
and other investigators have found that success in algebra is the lever we have as educators \cite{silver2008factors}. 
In this context, it makes sense to develop an approach to K-8 CS which will better prepare children to master algebra.

We call our approach ``algebraic thinking'', going back to the original use of ``al jabr'' (reunion of broken parts) by the Baghdadi mathematician al-Khwarizmi in the book which transmitted algebra to Europe \cite{aljabr2017oed}.
But, rather than factoring integers and polynomials,
we decompose complex shapes by identifying the constituent parts 
and teach children to build their own graphics up hierarchically.
By focusing on shapes,
children learn about recursive (tree) data structures by building increasingly complex pictures.
We implement this approach in 
GraphicSVG, a Domain Specific Language (DSL) embedded in the functional language Elm.  
Matching the semantics (and vocabulary) of drawing with stencils allows children to begin exploring
without stumbling over new concepts.  
Children start right away by modifying a list of shapes,
so initial learning is restricted to a short list of functions with small numbers of obvious arguments
and syntax for lists and for forward function application---pipelining---to better expose the combinatorial 
nature of shape construction and composition.
Children's ``inner scientist'' loves experimenting with possibilities
and rapidly assimilates the patterns.
Almost all language features can be introduced in a planned way after children are confident in their ability to handle programming.

\smallskip
Our project has developed the following tools to help us deliver this outreach program:
\begin{enumerate}
\item The DSL GraphicSVG for the algebraic construction of vector graphics, animations and interactive programs, embedded in Elm.
\item A web interface to the Elm compiler with limits designed to minimize beginner mistakes, and a chat system in which off-site mentors can see a child's code when they ask for help.
\item A web-based discovery tool.
\item Adaptable lesson plans for introducing drawing, animations, functions, variables and user interaction to children aged 10 to 14.
\item A prototype, iPad-based projectional editor for Elm with GraphicSVG.
\end{enumerate}
Based on the needs of our partners, we deliver workshops in different configurations, from one hour to one hour per week for seven weeks and ``hackathons'' of five hours.
We almost exclusively deliver our workshops to whole classes,
because we want to reach underrepresented groups
who are less likely to be enrolled in optional activities.  
Our school board partners help us target schools in high-needs neighbourhoods.
On a smaller scale, we also visit summer camps, and encourage summer camps to adopt our curriculum by hiring our volunteers, and we invite other individual students and groups to campus for drop-in sessions.
Most children are between 10 and 14 years old, with hackathon participants being 12 to 14.
Although sending undergraduate instructors to schools makes it difficult to scale up,
teachers really like to have role models in their classrooms.

\smallskip
This paper is organized as follows:  
In section \S\ref{sect:background}, we cover the background ideas which motivate our approach, some history of introducing programming in K-12 education, the importance of algebra in student success, and the evidence that using a functional language promotes success in algebra.
All of which motivates our adoption of the term Algebraic Thinking.  
We then explain the expected and previously observed benefits of Social Constructivism.  Finally we describe the important features of Elm relevant to this work. 
 
In section \S\ref{sect:design}, we explain the design of graphics and interaction library, and give two examples of how this design translates into particularly simple and attractive lesson plans.
In section \S\ref{sect:experience}, we describe our experience delivering these lesson plans in diverse schools and at different grade levels.
In section \S\ref{sect:related}, we discuss two related works in progress and works of other researchers.

Finally, in section \S\ref{sect:conclusion}, we conclude with a summary of what we have learned and how this will shape our future plans.

\section{Background}\label{sect:background}

It would be impossible to summarize the many threads of research and teaching practice attempting to integrate or leverage computers in education,
but most trace back to Papert and coworkers at MIT \cite{Papert:1980:MCC:1095592},
who were not primarily interested in teaching programming, or recruiting future software developers.  
They were interested in how children learn, and primarily interested in teaching them to discover mathematics for themselves by forming hypotheses, performing experiments and reasoning.

\vspace{-1em}
\subsection{Algebraic Thinking}
Why do we characterize our approach as ``algebraic thinking'', and not ``computational thinking'' or ``algebraic reasoning'' or ``algebraic program construction''?
The term  ``computational thinking'' was introduced by Wing \cite{Wing:2006:CT:1118178.1118215}
to draw attention to the need for increased enrolment in CS at all levels.
There are now many outreach programs aimed at K-12, including well-funded non-profit organizations,
however there are also some pointed questions being asked.
In particular, what is the difference between ``computational thinking'' of 2010 and ``programming'' of 1980, and the answer seems to be ``not very much''.
Literacy researchers have observed the same patterns (failure to define or agree on aims, cycles of reinvention) in other literacy movements \cite{vee2013understanding},
and it is a shame that we have not learned more from them.

Numeracy and more general mathematical ``literacy'' have their own history,
with a recent understanding of the role of algebra as a gateway to high school success \cite{silver2008factors},
and the importance of preparing students for high school algebra \cite{doi:10.3102/00028312040002353}.
Pre-algebra starts as early as kindergarten with patterning (e.g., red-red-blue-red-red-blue-?), and some researchers also use the term Algebraic Thinking to describe the basket of cognitive skills needed and promote methods of fostering them \cite{kieran2004algebraic}.

Recently, Guzdial \cite{doi:10.2200/S00684ED1V01Y201511HCI033} surveyed the literature on teaching computer science, focusing on K-12.  He is also critical of the simple definition of ``computation thinking'', and he discusses at length Soloway's Rainfall problem which highlights the consistent failure of conventional approaches to teach even undergraduates the basics of programming. 
In contrast, Fisler found in \cite{Fisler:2014:RRP:2632320.2632346} that students using functional languages ``made fewer errors than in prior Rainfall studies and used a diverse set of high-level composition structures''.
So there is some evidence that a functional approach is superior for beginning undergraduate and high-school learners.  Such an approach could be called ``functional thinking''.

\vspace{-1em}
\subsection{Social Constructivism}

Social constructivism is the theory of learning which focuses on peer interactions, 
especially the sharing of information and construction of shared explanations and procedures.
Socially constructive learning is thought to be better for acquiring higher-level skills,
but it can only work if children are engaged enough to stay on task (or, even better, invent their own tasks).
It is closely associated with Papert, 
but although far from novel, it is underexploited.
We have relied on it in our other outreach activities, which has inspired our attempts to use it for Elm programming.  We describe the two most important activities:  when teaching binary numbers, we encourage children to help each other out in using our iPad app \href{https://itunes.apple.com/app/image-2-bits/id967807383}{\underline{Image 2 Bits}}, in which students encode black and white images with binary numbers.  The students each create an image on their iPad; the iPads share the encoding of that image with other students, and the students then decode each other's images.  Once they are done decoding the image, students have the option to send a ``like'' to the student who created the image.  Similarly, when teaching a (simplified version of) Computed Aided Tomography (CAT scans), we rely on group reinforcement, motivated by a game.  Students arrange themselves in a grid to represent the body part being imaged.  They act out the x-rays passing through the body and partly being absorbed by bone and muscle tissue.  Each student represents either a piece of muscle or bone.  The x-rays pass directly through the muscle tissue, but get partly blocked by the bone tissue.  In order to correctly calculate where the bone and muscle tissues are, each student in the group must pay attention and correctly report how much light comes through.  Then the students work together to solve for the location of the bone and muscle.  Where the numbers allow, we have two or more groups work in parallel, competing to reconstruct their image first.  That games have rules does not surprise children, and they immediately realize that everyone must fully understand their role, or the whole effort will collapse.  Furthermore, they actively think about the algorithm they are applying in an effort to find ``shortcuts''.

Allowing children to consult each other and work together in groups can easily lead to off-topic chatter, rather than collaboration,
but we have found an easy way to gauge the effectiveness of unstructured learning:
If we observe that techniques explained to one group of students propagate across the classroom, then we know that chatter is actually productive, and that we have successfully adjusted our goals and teaching style to the class.

\vspace{-1em}
\subsection{Elm}
Elm (http://elm-lang.org) was imagined as a vehicle for delivering best practices in program language design to web front-end developers \cite{czaplicki2012elm}.  It is deliberately simple, for example, dropping support for Functional Reactive Programming in version 0.17 \cite{farewell-to-frp}.  
It looks like other ML-derived languages, with algebraic datatypes, but it does not have user-defined type classes.
For beginners, this is a good balance between expressiveness and simplicity (particularly of type errors).
Languages with record types, but lacking fully algebraic types, require encoding of state into strings and other error-prone idioms which place additional burdens on the programmer.
For example, the Elm compiler will not accept incomplete case expressions.
It has a foreign-function mechanism designed with safety first.
When it was initially developed, it was the natural language for Haskell programmers needing to create a JavaScript app, and was a radical simplification of the complexity of JavaScript, HTML and CSS.  Today, such developers have many other options including PureScript (http://www.purescript.org), TypeScript (https://www.typescriptlang.org/), or Flow (https://flow.org/).

	In Elm 0.17, Functional Reactive Programming features were removed in favour of The Elm Architecture (TEA) \cite{farewell-to-frp} with subscriptions. It has three parts, which in our case are typed as
\begin{itemize}\setlength{\itemsep}{-1pt}
\item a \texttt{model} data type for state, with an initial value, 
\item \texttt{view : model -> Collage (Msg userMsg)}, and
\item \texttt{update : userMsg -> model -> model}.
\end{itemize}
	
The \texttt{model} is a data type which contains information about the current state of the program. For example, the model might contain information about the current time and what slide you're on in the case of a presentation.
	
The \texttt{view} function visually represents the state stored in the \texttt{model}, usually through HTML, or, in the case of GraphicSVG, through SVG. Elm's functional paradigm eliminates many pitfalls found in traditional web development, as it guarantees that the same state always produces the same view.
	
The \texttt{update} function takes both the \texttt{model} and a data type (known as the ``Message") as a parameter. The \texttt{update} function implements all transition functions from one state of the program to the next. Some messages, such as time updates, are requested from the Elm runtime, while special transformers that can be applied to any Shape tell the run-time to generate specific messages based on user interaction.  
Elm's runtime redraws automatically based on need.

Throughout the text, we use \texttt{userMsg} for the type variable which is always the type of the messages the run-time system generates and passes to the \texttt{update} function.
We need to use a type variable, because the number and type of messages will change 
as the user adds buttons with a \texttt{notifyTap} transformer, 
or requests tap location within a \texttt{Shape} using 
\begin{verbatim}
notifyTapAt : (( Float, Float ) -> userMsg) 
              -> Shape (Msg userMsg) 
              -> Shape (Msg userMsg)
\end{verbatim}
and similar functions.
This is an example of where the stronger typing available in Elm is used.
We do not yet know how to explain type parameters to children of this age,
but completely hiding them would require separate libraries for drawing interactive and non-interactive graphics.

\smallskip

As a teaching language, Elm offers simplicity in both syntax and semantics,
and because it compiles to JavaScript, it is accessible in schools without installing software.
It works on every platform with a web browser,
and it is easy to build collaboration and distance-learning tools around it.

\section{Instructional and Tool Design}\label{sect:design}

Our instructional design is tightly linked to our tool design.
At the core is our library GraphicSVG, \S\ref{sect:GraphicSVG}, which is a DSL for the algebraic construction of shapes, and upon reflection, it is really the language students start learning,
to the extent that the concept of variables, \S\ref{sect:VariablesFunctions}, is either delayed until the second workshop, or introduced to individual students whose graphics are complex enough to warrant them, 
and similarly for the concept of functions.
In the next subsection, \S\ref{sect:ElmEmbedding}, we review our language requirements, and how Elm meets them.
We go back to instructional design in \S\ref{sect:Interaction}, where we explain how we introduce state diagrams 
and then translate the concept into Elm code.
Finally, in \S\ref{sect:chat}, we describe our facility for distance mentoring.

\vspace{-1em}
\subsection{GraphicSVG}\label{sect:GraphicSVG}
\begin{figure}[htbp] 
   \centering
   \includegraphics[width=4in]{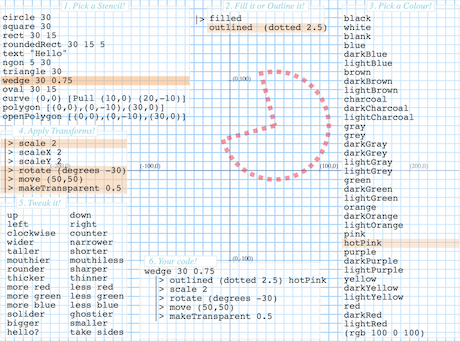} 
   \caption{ShapeCreator:  As a consequence of GraphicSVG's design, 
                 we were able to expose the combinatorial nature of shape construction
                 in an interactive tool for API discovery. 
                 The tool is presented as a menu in which the user can pick one Stencil in section 1; either the \texttt{filled} or \texttt{outlined} functions along with optional line style, and required colours, in sections 2 and 3; any number of transformations in section 4; and given options to adjust arguments to all of the above functions in section 5.}
   \label{fig:ShapeCreator}
\end{figure}

Our Graphics library, GraphicSVG, is based on the original Elm Graphics module which targeted HTML canvas elements, and it is partially backwards compatible.  GraphicSVG's principal types (\texttt{Stencil}, \texttt{Shape} and \texttt{Collage}) model real-world concepts: \texttt{Stencil} describes a recipe for creating a shape; for example, a circle with a certain radius, a rectangle with a width and height or text with a certain font and size:

\begin{verbatim}
circle, square, triangle: Float -> Stencil
rect, oval: Float -> Float -> Stencil
roundedRect: Float -> Float -> Float -> Stencil
text: String -> Stencil
\end{verbatim}

But, like a real-life stencil, a visible shape is not created until the user fills it in or traces its edge:

\begin{verbatim}
filled: Color -> Stencil -> Shape userMsg
outlined: LineType -> Stencil -> Shape userMsg
\end{verbatim}
Thus, a concrete analogy  explains why shapes cannot show up on the screen unless they are filled or outlined. 
This architecture limits the number of parameters each function takes,
making them easy to learn or even guess,
and the types match students intuition closely enough that we do not have to talk about them.
The limited number of arguments also make it easy to put all the basic functions
in an interactive crib sheet (Figure~\ref{fig:ShapeCreator}).
The types \texttt{Stencil} and \texttt{Shape} are introduced to students orally 
and we use the structure of the ShapeCreator to reinforce this.
We do not use type signatures, but students see the type names in compiler errors,
e.g., when they try to move a \texttt{Stencil} rather than a \texttt{Shape}.
They never use constructors for these types directly,
using exposed functions instead, some of which simplify the underlying type construction.
Fortunately, Elm's type errors (e.g. found a \texttt{Stencil} where a \texttt{Shape} was expected)
also match their intuitive understanding of these types, and need little explanation,
and, so far, students who choose to attempt more complicated user interaction 
are able to build a workable understanding of Elm types from there.
(See the last section to learn how we plan to use types as a teaching tool in the future.) 

\subsubsection{Design Motivation: Instructional Scaffolding}
	One aspect of Elm's original Graphics library that we found successful in classrooms was its support for instructional scaffolding. \cite{TRTR:TRTR68} The instructor shows students how to draw basic shapes on the centre of the collage. The next inquiry most students have is a logical progression from the current state: ``How can we move the shape from the centre of the collage?'' Like with \texttt{filled}, we use forward function application (\texttt{|>})\footnote{Elm provides two function application operators,
	\texttt{<|} is like Haskell's \texttt{\$} and  \texttt{|>} flips the argument order.} to apply transformations to \texttt{Shape}s,
	thereby visually laying out the combinatorial nature of shape construction. 

These functions have type (parameters)~\texttt{-> Shape userMsg -> Shape userMsg}:

\begin{verbatim}
move: (Float, Float) -> Shape userMsg -> Shape userMsg
scale: Float -> Shape userMsg -> Shape userMsg
rotate: Float -> Shape userMsg -> Shape userMsg
\end{verbatim}

	There are several advantages of this approach. First and foremost, it allows a very fast startup. By separating transformations from the shape itself, within a minute or two, the instructor can create a shape on screen simply by defining the bare minimum amount of information; for example, a radius and a colour is all that is needed for a circle, which follows the students' expectations about how to represent shapes.  It is also easier to remember parameter order than for similar functions in other languages which either have multiple parameters, confounding size and position, or use stateful drawing models. 
	
	For many students, this is their first exposure to a ``real'' programming language---or any programming language---and as such, seeing text produce a shape on screen is very exciting, and encourages them to ask questions which can lead the rest of the presentation. 

\subsubsection{GraphicSVG Apps}
	GraphicSVG contains three types of ``apps'' graded by complexity. 
	
	The first and most basic one, \texttt{graphicsApp}, allows the static drawing of graphics on the screen. It can be easily understood as requiring only the ``view'' portion of The Elm Architecture's model-view-update architecture, hiding the ideas of model and update from the student.
	
An intermediate app, \texttt{notificationsApp} allows the user to add interaction to their graphics
by adding notify transformers which cause messages to be sent to the student's update function when, for example, a shape is clicked:

\begin{verbatim}
notifyTap: msg -> Shape userMsg -> Shape userMsg
notifyEnter: msg -> Shape userMsg -> Shape userMsg
notifyLeave: msg -> Shape userMsg -> Shape userMsg
\end{verbatim}
See \S~\ref{sect:Interaction} for an example using notificationsApp to add interaction to a GraphicSVG application.
	
	Finally, \texttt{gameApp} provides the functionality of \texttt{notificationsApp}, but also has a parameter for a special kind of \texttt{Tick} message type, which, on each frame, will send their \texttt{update} function the time in seconds since the app was started as well as information about keyboard presses. Given that \texttt{gameApp} has only thus far been used for sporadic half-day workshops, few students have thus far taken advantage of the advanced features provided by \texttt{notificationsApp} and \texttt{gameApp}, other than support for animation. However, \texttt{gameApp} has been used internally to develop game templates used for day-long Hackathons where students compete in teams to create educational games, which have been very well received.

\vspace{-1em}
\subsection{Variables and Functions: a tool for code reuse}\label{sect:VariablesFunctions}
Graphics can be used as an effective way for students to learn the fundamental concepts behind mathematical functions. After a brief introduction of the ``input-process-output'' model, we asked a class of Grade 7/8 students to come up with real-world examples of processes which can be described using this model.  They shared a wide variety of ideas: printers (inputs: blank paper, ink, information, energy; pro\-cess: place ink on to paper in the correct pattern; output: printed paper), schoolwork (inputs: constraints, objectives, rubrics; process: completing the work according to necessary steps; output: completed assignment/presentation) and, somewhat comically, paperwork (inputs: forms to fill out; process: collect and write down information; output: completed paperwork), among others.  The mentor stressed that almost any real-world process can be generalized using this idea.

Directly following the discussion, we used the example of the flower  (Figure~\ref{fig:Flowers}).  The mentor demonstrated how defining a separate variable with a GraphicSVG group,
\begin{verbatim}group: List (Shape userMsg) -> Shape userMsg ,\end{verbatim}
can be used to easily create copies of a graphic on the collage plane.  However, this time we presented a new motivation: how can we create copies of the flower which are different colours?  Once again, students suggested the obvious ``copy and paste'' approach, with \texttt{flowerRed}, \texttt{flowerGreen}, \texttt{flowerBlue}, etc. variables being created and then referenced in their collage.  
The volunteer instructor then emphasized the point that similar code should be reused, instead of duplicated. The ultimate motivation then becomes that learning how functions apply to Elm would benefit their more rapid and straightforward creation of new artwork and animations. Thus, the concept of functions was presented as a ``tip'' rather than a traditional lesson.
	The instructor then demonstrated how to add an input to the flower function, a term with which they were familiar due to the aforementioned discussion:

\begin{figure}[htbp] 
   \centering
   \begin{tabular}{m{0.2in} m{5.5in} m{5.5in}}
    a) & \includegraphics[width=5.5in]{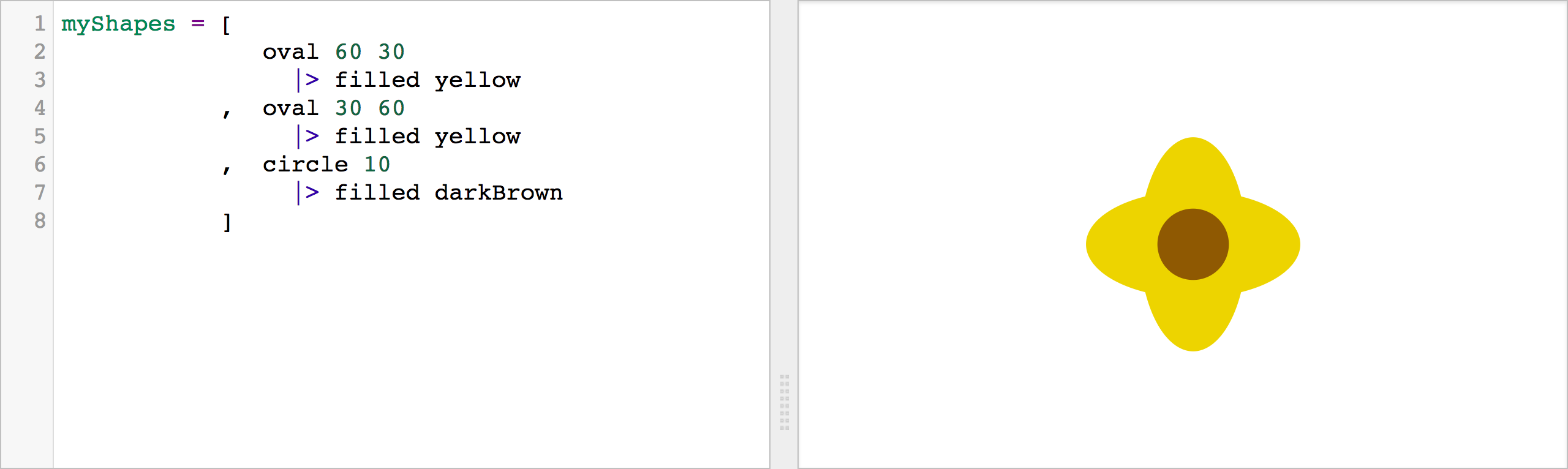} \\
    b) & \includegraphics[width=5.5in]{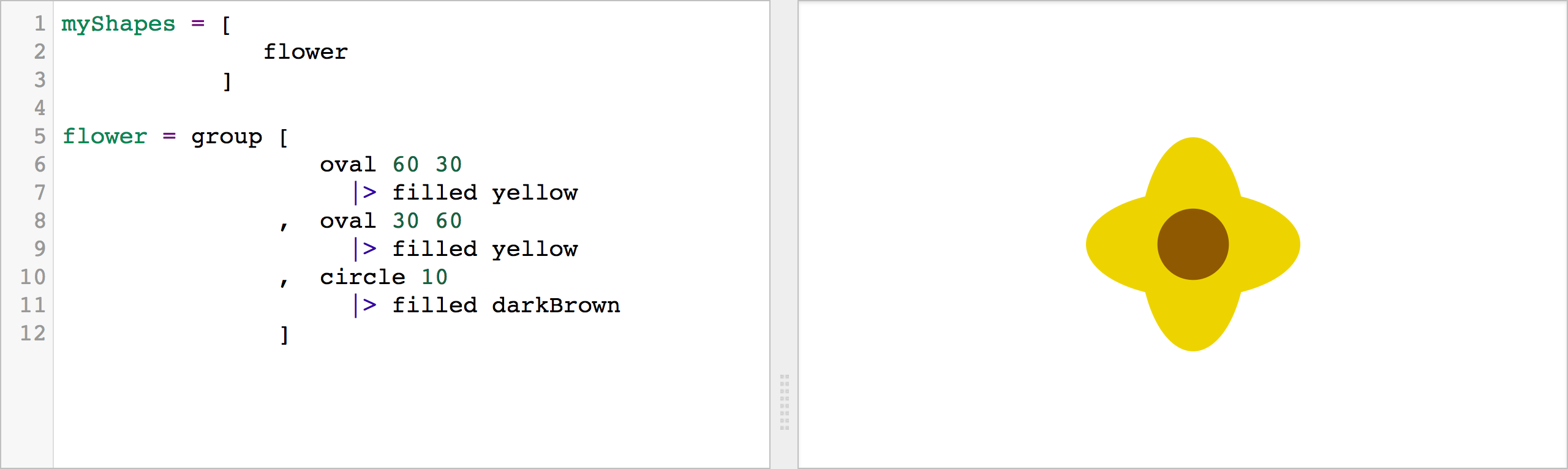} \\
    c) & \includegraphics[width=5.5in]{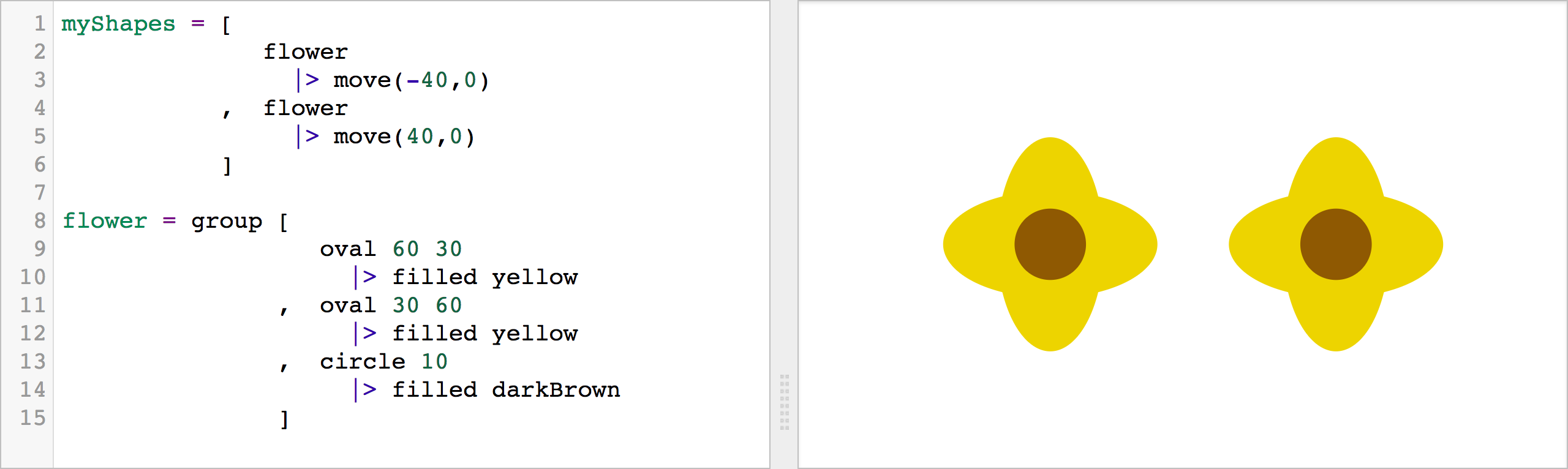} \\
    d) & \includegraphics[width=5.5in]{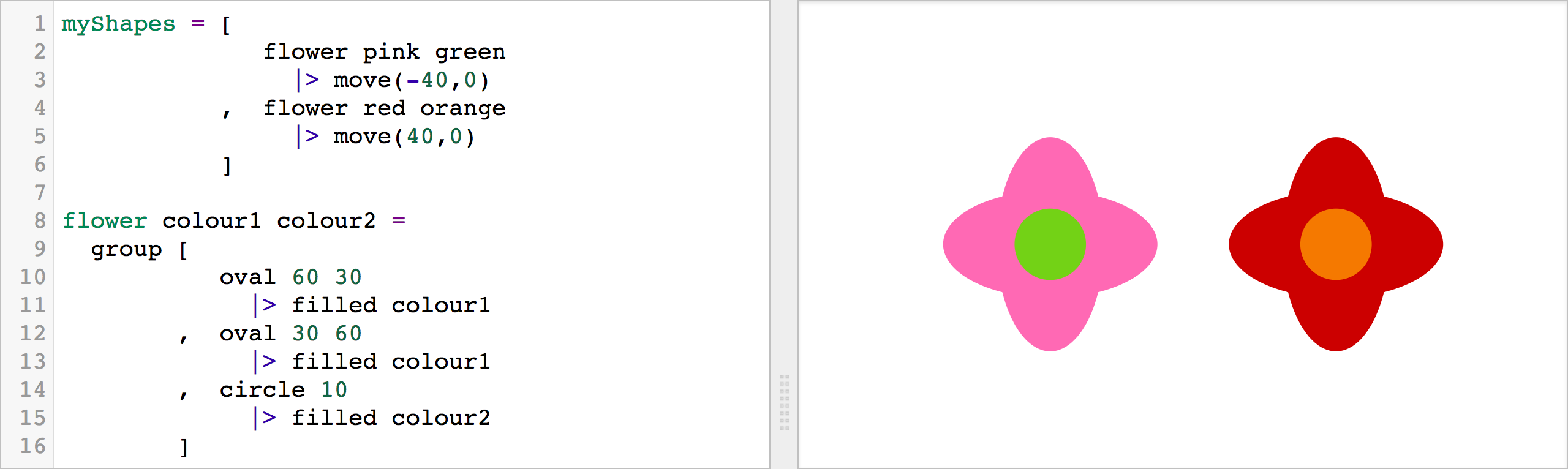} 
  \end{tabular}
   \caption{Introducing variables and functions as progressively more powerful ways of reusing code.
   Initially in a) there is one flower, and without instruction, children will copy and paste the shapes to produce multiple flowers, but in b) they are shown that the ``flower'' can be given a name \texttt{flower} to make the purpose of the shapes clear, and so that in c) it can be used multiple times.  Finally in d) it can be transformed into a function to create less boring flowers.}
   \label{fig:Flowers}
\end{figure}

\begin{verbatim}flower colour = group [...]\end{verbatim}
upon receiving a compiler error on the projection screen, students were then able to successfully determine that the colour would need to be added after the name of the shape in the \texttt{myShapes} collage:

\begin{verbatim}
myShapes = group [
                      flower green
                           |> move(-50,0)
                 ,    flower blue
                           |> move(50,0)
                 ]
\end{verbatim}

Upon compiling the program, the students then found that the flowers had not yet changed colour. After some discussion, they were able to determine that the \texttt{colour} variable must be used in lieu of a specific colour:

\begin{verbatim}[oval 50 30 |> filled colour]\end{verbatim}

After this guided example, students were able to independently come up with a strategy when asked: what if we want to be able to set an arbitrary colour for the centre of the flower? After a subtle hint of adding a ``1'' to the colour input, becoming \texttt{colour1}, students were able to leverage their knowledge from creating a single-input function and solve the same kinds of problems that came up in the single-input example, which allowed them to describe to the mentor how to implement a function with two variables using Elm (Figure~\ref{fig:Flowers} d). 
	Thus, students were able to map the ``input-process-output'' concept to three different realms: an abstract concept, describing real-world processes, and, finally, implementing the idea algebraically by leveraging graphics programming with GraphicSVG. Many students made the connection to their math classes where they had begun to talk about mathematical functions and lines but were motivated by being able to use the concept to improve their Elm graphics. Future work would include devising methods to determine quantitatively if the students who learned functions in terms of Elm programming have a more concrete idea of the concept than students who learned it only mathematically.

\vspace{-1em}
\subsection{Embedding in Elm}\label{sect:ElmEmbedding}
We have already discussed the use of Elm in different places,
but with the previous context, we can now better explain that we require a language with
\begin{itemize}\setlength{\itemsep}{-1pt}
  \item pure functions matching the child's inherent idea of tools, like stencils, pens and paint brushes,
  \item facility for creating a domain-specific language,
  \item strong types with simple type errors, like ``Stencil expected, but found a Shape'',
  \item minimal syntax to learn, and
  \item a minimum of matching delimiters,
\end{itemize}
serves language designers and children alike.
We could have designed a language from scratch,
but that would have been more work.
Choosing a language which compiles to Javascript makes it easy for children to share their accomplishments with friends and family,
and to eventually graduate from programming in the GraphicSVG sandbox to more general web programming.
Or, because Elm is so close to Haskell, they could go on to learn about type classes, and not run out of material for a long time.

Although the minimization of matching delimiters and things like the strictness of rules on leading whitespace may seem to be minor points, when children encounter these issues,
they really distract from the important concepts they are supposed to be learning.
In particular, the \texttt{|>} infix operation, which is defined by 
\begin{verbatim}
  x |> f  ===  f x
\end{verbatim}
allows us to order the functions involved in shape creation in the natural order in which we introduce them, and in which they are ordered in in the ShapeCreator (Figure~\ref{fig:ShapeCreator}).
	
\vspace{-1em}
\subsection{Interaction} \label{sect:Interaction}
\begin{figure}[htbp] 
   \centering
   \includegraphics[width=2.7in]{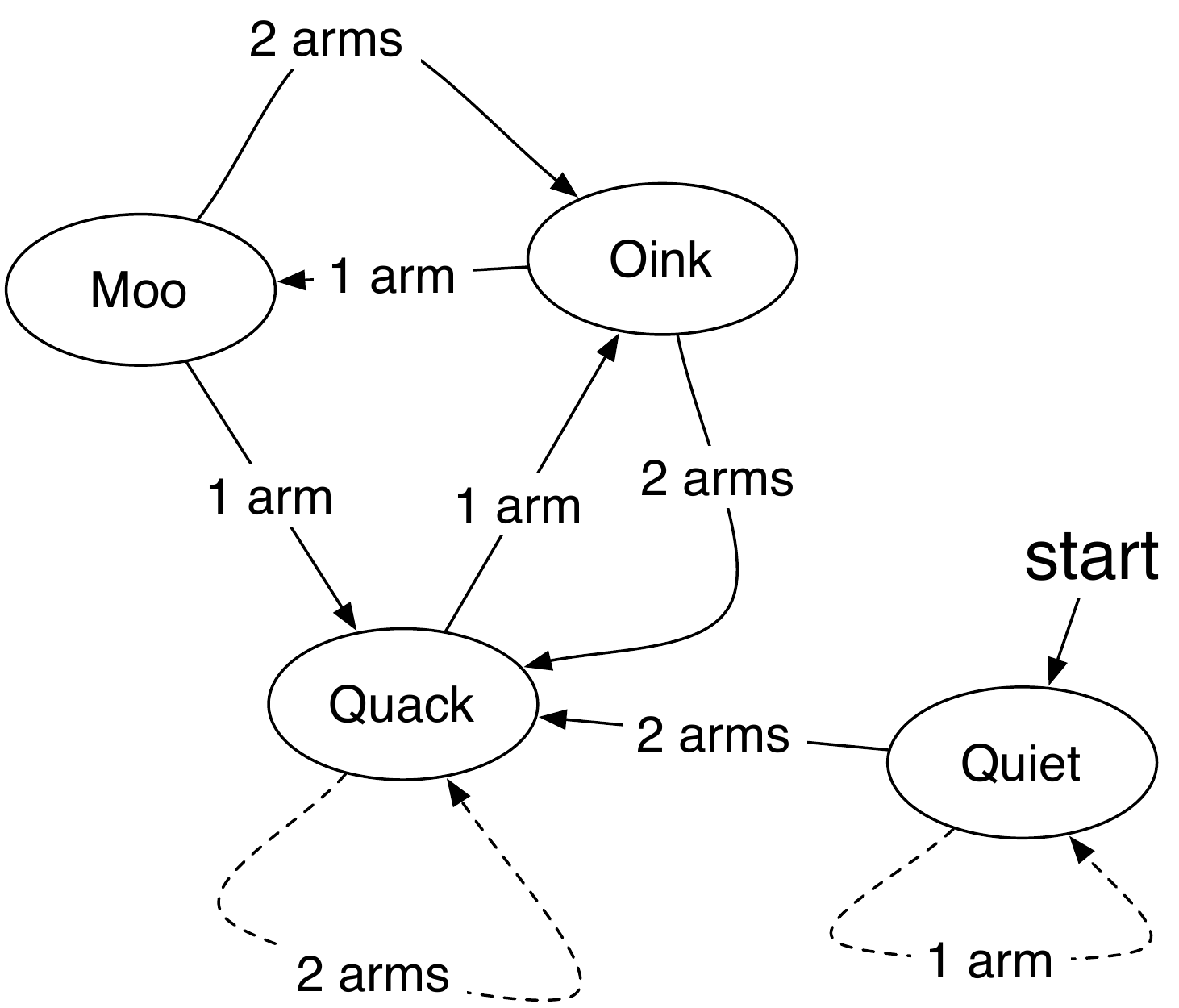} 
   \caption{Without explanation, students are given a state diagram similar to this one, and an instructor waving arms (or clapping, flashing lights, etc.).  It doesn't take long for them to understand what state is.}
   \label{fig:MooOinkQuack}
\end{figure}

Although the Elm Architecture depends on separating state, updates and views,
we do not explain state until we are ready to introduce interaction.
Although we use different types of interaction in our learning tools,
we only teach touch or click interactions.
Any object can act as a button by simply adding the \texttt{notifyTap} transformer
in the same way that they add spatial transformation.
So, unlike with most languages,  syntax is a very low barrier,
and the essential barrier is understanding state and transformations.
To make it memorable, we introduce state by putting up a simple
state diagram (Figure~\ref{fig:MooOinkQuack}) and acting it out until the
whole class understands the game.
Children are very attuned to games, and understand the need to learn the rules,
so by making a State Diagram into a game, they control their own learning.

\begin{figure}[htbp] 
   \centering
   \includegraphics[width=5.5in]{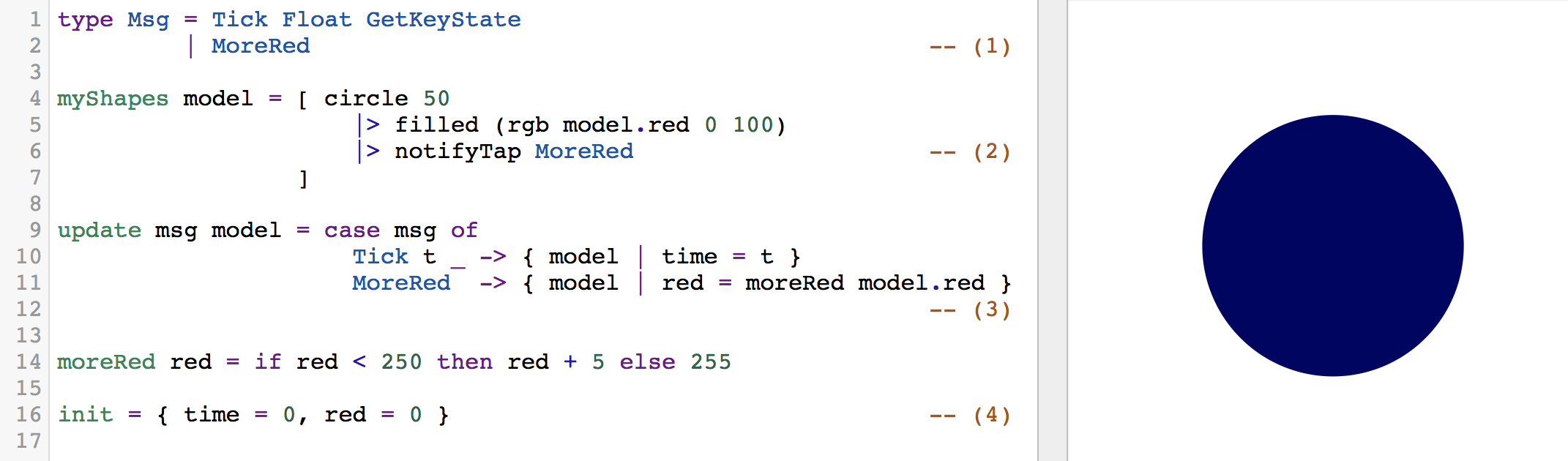} 
   \caption{When students are familiar with animations using model.time, it is relatively painless to add user interaction by (1) adding messages to the \texttt{Msg} type, 
   (2) adding a notify transformer to any shape, (3) adding a case to the update function, 
   and (4) adding initial values to any new state components.}
   \label{fig:MoreRed}
\end{figure}

Once they have understood states and transitions,
we can show them how to translate the state diagram into Elm by
turning each state into a constructor for a state data type,
and each transition function into a function.
We have only taught interaction to self-selected groups of children,
so we have been able to jump from the translation of one diagram
to the identification of other state diagrams they encounter in real life,
and then state as it exists in games.
After that, we show them a simple example like Figure~\ref{fig:MoreRed},
and point out that the time they have been using in their animations is also
as state which was updated for them in code they were ignoring from the game template.
There are only four steps to adding support for interaction:
\begin{enumerate}\setlength{\itemsep}{-1pt}
   \item adding messages to a user-defined message type,
   \item adding to any ``buttons'' a notification transformer with the appropriate message,
   \item adding to the update function, possibly calling simple state transformation functions, and
   \item adding initial values for any new state components.
\end{enumerate}
Of course they can make arbitrarily complex use of state in their existing views.

State was the subject of lively discussion and even controversy at the conference,
perhaps because it is often associated with error-prone use of global variables in imperative programs.
In fact, we would argue that many tools relied on by functional programmers,
including monads and functional reactive programming,
are syntactic sugar designed to make interacting with complex states safe.
It is easy to think that exposing state goes against the philosophy of functional programming.
We view state as an inherent property of interactive programs,
and it is best to make it explicit so we can talk about it,
rather than hide it using abstraction or syntactic sugar.
In fact, the successful Bootstrap program makes similar use of explicit state\cite{BootstrapReactiveUnit2}.

\vspace{-1em}
\subsection{Distance Mentoring}\label{sect:chat}
\begin{figure}[htbp] 
   \centering
   \includegraphics[width=4in]{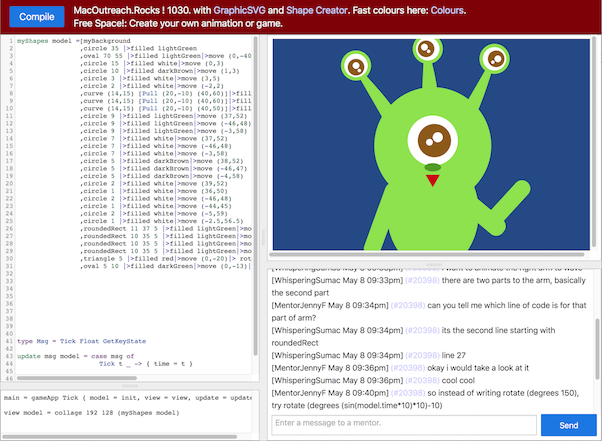} 
   \caption{Students are given unique identifiers and can send questions to mentors, which will give mentors access to the student's code. The mentor can reply with suggestions or fixes, and even attach working code that the student can manually incorporate into their version.}
   \label{fig:GreenAlien}
\end{figure}

\begin{figure}[htbp] 
   \centering
   \includegraphics[width=5in]{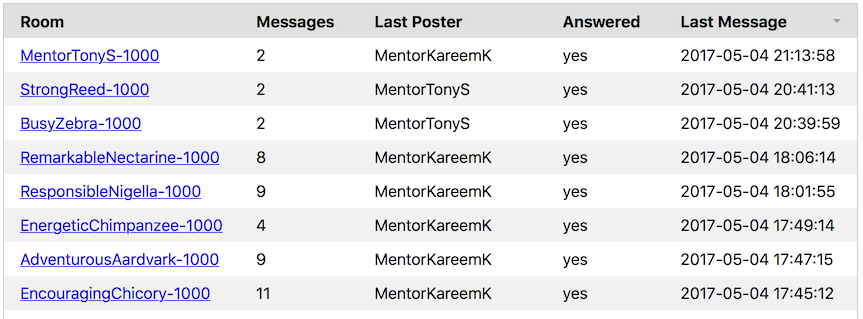} 
   \caption{Mentors have access to a sortable list of questions posed by students as part of our new distance learning approach.}
   \label{fig:MentorListView}
\end{figure}

We have recently added an online mentoring system to our web programming environment. 
Each student signs in and picks  a ``game slot'' to work in.
Some slots have challenges with specific goals, like adding a new ice cream flavour to a vending machine,
but most students opt to create pictures or animations in one of the free-form slots,
or one of the slots containing a game templates.
Each slot has a chat, accessible only by that student in that slot (Figure~\ref{fig:GreenAlien}).
Mentors have access to a view of all active discussions (Figure~\ref{fig:MentorListView}) 
which indicate which discussions have unanswered questions.

The discussions are presented as ``help lines'' where students can ask any questions about their code (working or broken), 
and mentors can send back answers or even links to differences between the student's code and code modified by the mentor.
During scheduled workshops, we make sure someone is monitoring the discussions and answering questions immediately.
The mentor's modifications are to a private copy, not the student's code, or the code of another mentor.

This system is new, but students are already using it to ask questions after school hours, as well as to get answers faster during lab time when the on-site instructors are busy helping another student.
We anticipate using it to support teachers who have received training, 
but who would not attempt to teach programming without such backup support.
We have two teachers who plan to pilot this usage.

\section{Experience}\label{sect:experience}

One of the advantages of focusing on graphics so early,
is that it allows every student or small group of students to create their own image,
which gives them an early sense of accomplishment.
Many teachers comment, in particular, on the engagement of children who very rarely engage with conventional mathematics instruction.
Several of our undergraduate instructors who did not excel in mathematics at that age have commented that a class learns pretty quickly which students will always answer questions first, and this can be discouraging.
Designing a personal or small-group graphic, however, \emph{cannot} be left to the ``math geniuses''.  
Our Hall of Fame\footnote{\url{http://www.cas.mcmaster.ca/~anand/hall.html}} shows that there is a lot of creativity in students this age.

Working in pairs works best for most students,
and since each team has their own project, interactions between teams is positive
and focused on sharing techniques.
We encourage students to share knowledge and achievements,
and present more advanced topics as tips they may find useful for their current creation.
With previous activity designs and other programming languages, we frequently had to devise strategies to keep children on task, including the interleaving of girls and boys to keep friends from engaging in off-topic discussions.
However, after switching to GraphicSVG/Elm, we can encourage children to consult each other.
Now, if one team asks for help to make their shape blink, we will soon see half the class following suit in their animations.

Younger children (less than 10 years old) do still stumble over syntax, probably because they are just discovering punctuation in English, so they cannot build on that knowledge.  
 For these students we will describe ongoing work to create an error-free editor just for them.
 
Children have no trouble incorporating time as a variable any place a number can be used,
which allows them to animate their graphical creations.
Once they start repeating elements, we introduce variables,
which may expose that they do not understand that the view function has the current state as an input, and therefore access to time.
When this arises, we explain how to fix this by adding a time argument to abstracted functions,
but we do not spend enough time with most children for them to encounter this problem, or for them to understand how to plumb arguments through nested functions.
Once they ask about user interaction, we introduce state.

None of our students knows trigonometry, so we give them an example using sine to make an object move back and forth, usually after the first group discovers that 
\begin{verbatim}
    move (model.time,0)
\end{verbatim}
will move their shape off the screen, never to return.
Since the conference in the summer, we have added multiple tabs to the ShapeCreator (Figure~\ref{fig:ShapeCreator}) including a tab devoted to applying sine and cosine functions to translation, rotation, scaling and colour changes.

Approximately 150 children who had attended workshops in previous years were taught how to add tap and click detection to their programs,
although they were not required to add interaction, and were encouraged to continue work on animations if they were still working on that.
Of those children about half chose to experiment with interaction
by modifying the example code written by the instructor,
in which a single clickable rectangle moved around the screen, 
and when clicked would update a score and change a state.
Because any Shape, even invisible Shapes can detect clicks, most children
incorporated their previous Shapes (especially Shapes with faces, including an animated turnip!) as buttons,
and found it quite rewarding to make difficult games in which the clickable object moves, grows and shrinks in unpredictable ways.
A handful of students added multiple click targets, e.g., 
by making the nose and body clickable, adding points for clicking the nose,
but subtracting points and displaying a message when clicking on the body.
State Diagrams give children a concrete representation to reason about something which they immediately recognize as a concept underlying the implementation of games.  Even simple games such as Snakes and Ladders involve recognizable states and transitions, and most video games have levels and power-ups.  
Anecdotally, all children seem to find concrete State Diagrams useful,
even children with the highest levels of confidence and experience,
who have an ``aha'' moment when they realize that this unifies a lot of their experiences.

\vspace{-1em}
\subsection{Adoption}
Teachers are busy, and the curriculum is already full enough.  
One indirect way in which advancing technologies are indirectly making teaching even more challenging is a reaction to the expected displacement of workers by automation  \cite{Frey2017254}.
In this context, teachers and schools in our area are trying to develop creativity and teamwork (referred to as 21st-century skills),
which further squeezes time for pre-Computer Science.
In this context, what we need is a method of introducing Computer Science that
provides at least as much progress toward existing curricular goals as 
the instruction it will displace.
We have relied on teachers' comments to judge how well we are meeting this goal,
as we rely on peer recommendations to be invited to new classrooms. 
Two quotes from teachers whose classes we visited in June:
\begin{quote}
This morning's coding session was fantastic!  Love that I see math curriculum connections with integers and placing co-ordinates on a grid!
\end{quote}
So, the connection with coordinates meets teachers' needs to cover curriculum, and their students are so excited by their ability to independently explore the material, that (at least anecdotally) their engagement spills over into other subject areas.
\begin{quote}
My students were still talking about [the workshop] last week over the last few days so it
certainly had an impact on their learning and engagement!
\end{quote}
So, although we see the acquisition of algebraic thinking as the bigger and more important target,
the visible gains in enthusiasm for geometry easily justifies the time taken,
while we figure out how to measure the impact on algebra.

\section{Related Work}\label{sect:related}

As a rule, tools meant for experts are not suitable for learners.
This is certainly true of integrated development environments with lots of key-press-saving state.
Many practitioners who see functional programming as the reserve of elite programmers have asked us why we use it for beginners.
In this case, functional programming's fundamental advantages actually play out in the favour of beginners.
Pure functions are much easier for beginners to reason about.
Declarative ``variables'' match preexisting expectations from algebra,
and will not cause confusion for students who have yet to learn algebra.
They are readily accepted as being shortcuts to avoid typing the same thing over and over.
And finally, the surfacing of structure in functional programs is of benefit 
to designers at all experience levels.
Hughes' observations apply equally to beginners \cite{doi:10.1093/comjnl/32.2.98},
although they need to be incorporated into the instructional design,
creating a new shared experience for these students,
because it cannot be described relative to previous practices which have no meaning to beginners.
While we share a lot in motivation and methodology with the Bootstrap program, as articulated in \cite{Felleisen:2009:VWC:1538788.1538803},
especially the importance of functions,
we differ in some practical ways.
We see typing as the most effective way of boosting program quality, and something to start teaching early.
We find the concrete syntax of Elm much closer to school algebra than Scheme,
lowering the barrier to knowledge transfer.
Our building blocks stencils, shapes, groups thereof, and geometric transformers,
have a richer algebraic structure than the images used in Bootstrap.
Perhaps our younger audience finds this more natural, since they still spend a lot of time drawing and composing artwork from simpler objects.

Our focus on the algebra of shapes grew over time, starting with the use of Gloss\cite{gloss} with older children,
and subsequently the (now deprecated) Elm graphics library.
The ``many concise transformers'' design parallels Walck's approach in supporting learning through exploration\cite{DBLP:journals/corr/Walck16}.
Although we have created a Domain Specific Language for specifying diagrams,
ours does not support more advanced typing or layout engines,
as, e.g., \cite{Achten:2014:TOP:2746325.2746329}, \cite{Yorgey:2012:MTV:2364506.2364520},
because the result would not be as learnable in small chunks, 
and by having children do layout themselves, 
we get them to integrate visual art with mathematics.

\vspace{-1em}
\subsection{Related Project:  MacVenture}
In MacVenture \cite{Helen-Brown-thesis}, students create their own gamebook with nodes (places) and edges (ways).  This game arose out of a desire to make graph structures interesting to children.  Each place has a textual description, including textual keys, and ways can be annotated with matching locks.  Currently, students also choose images for each place, but this frequently causes network problems, and makes sharing (as demanded by students and teachers) impractical.  Incorporating touch editing for Elm graphics to replace the images will solve these problems.  

\vspace{-1em}
\subsection{Related Project:  Touch Editing}
We have done workshops where some students have written Elm programs on iPads and productivity is obviously lower.
All of the limitations of tablets are highlighted, and none of the advantages.
Given the prevalence of iPads in younger grades---with some schools phasing out desktop and laptop computers entirely---we knew we had to do better.
There are many examples to guide us:
successful non-textual editors (Scratch, Hopscotch, Lightbot, etc.);
early experiments in user-directed program transforming editors:  ``by precluding the creation of syntactically incorrect files, the Synthesizer lets the user focus on the
intellectually challenging aspects of programming" \cite{Teitelbaum:1981:CPS:358746.358755};
 ``autocompletion'' in contemporary IDEs; 
 and current research on projectional editors \cite{voelter2014towards}, \cite{omar2017hazelnut}.
\begin{figure}[htbp] 
   \centering
   \includegraphics[width=3.4in]{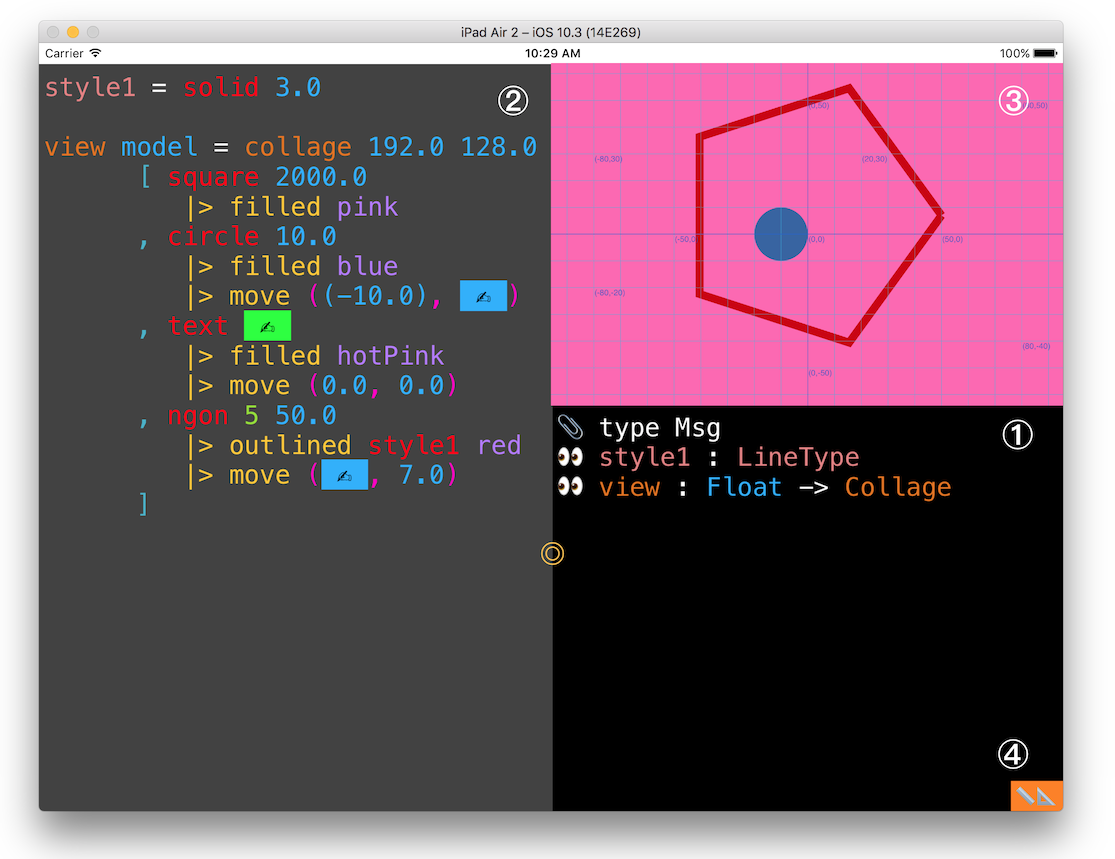} 
   \caption{The prototype app consists of three resizable panes: (1) is a list of definitions in the user's code, the visibility of which can be toggled on or off. In the main editor pane, (2), code is highlighted according to type. Tapping different elements brings up context-specific panes that allow the insertion of syntactically correct code (Figure~\ref{fig:InsertView}). The top-right pane (3) is the output generate by the user's code. Finally, the buttons at (4) allow the student to toggle helpful tools such as the Cartesian grid and a bezier curve helper. }
   \label{fig:ELMEdit}
\end{figure}

\begin{figure}[htbp] 
   \centering
   \includegraphics[width=3.4in]{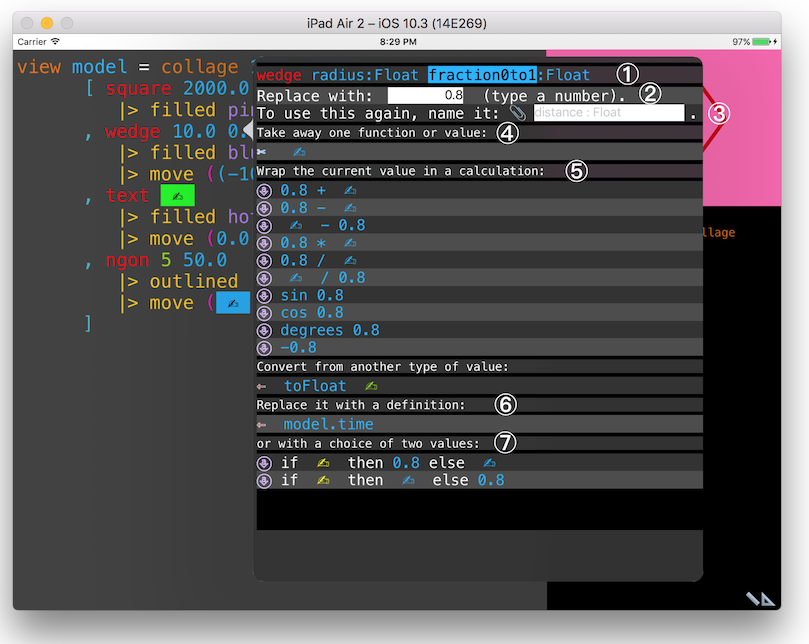} 
   \caption{Context- and Type-Aware Transformations: After tapping on an element, a popup gives the user options to modify the value.  The signature of the function using the element and position of the selected argument is shown at the top (1).  The user can type in a constant in (2), or make a new definition out of the selected value (3).  This is the only time that the user types the variable name, so there are no concerns about misspelled variables.  There is then the option to remove the current element (for example, an element from a list, or a transformer application), or replace the value with a hole (4).  Calculations incorporating the current value follow (5), as well as the definitions which already are in the context and match types (6).  In this example, there is only \texttt{model.time}, which is an record element of the \texttt{model} argument, but there could also be user-defined variables (3).  Finally, there are conditional options (7).}
   \label{fig:InsertView}
\end{figure}

In rethinking the division of labour between programmer and computer,
it seems obvious that there is no advantage in providing hints to the programmer
which are later found by the compiler to cause type errors.
To avoid type errors,
instead of editing text,
our users incrementally edit typed Abstract Syntax Trees (ASTs),
via touch-based interaction with the textual representation of the AST.
Interesting points about our implementation:
\begin{itemize}\setlength{\itemsep}{-1pt}
  \item Syntax highlighting helps readability and guides the user to create syntactically correct programs.
        Our programs are always correct and typed, so we can use colour for type, with a unique colour for every type. Holes are typed, and have default values so partially complete programs can be visualized (Figure~\ref{fig:ELMEdit}).
  \item Program transformations are supplied by the ``editor''.  This is the hard part.  While strong typing restricts substitutions of values and fully applied functions to reasonable numbers (when compared to the tens and hundreds of autocompletions offered by conventional editors), beginners need more structure so they can learn in layers.  We are working out the presentation by trial and error as we support increasingly complicated transformations (Figure~\ref{fig:InsertView}).
\end{itemize}
This editor is still in development, but we have tested it with a dozen classes from ages 7 to 12, and early feedback indicates that it will make the initial steps easier.  The youngest group, a split grade 2/3 class (ages 7 and 8) were productively creating graphics with moving shapes within half an hour.  Some grade 7/8 students were able to effectively use variables within an hour.

\section{Conclusion and Reflection}\label{sect:conclusion}

Although our workshops initially targeted grades 7 and 8 (ages 12-14), 
we found early on that volunteer instructors preferred teaching younger children (aged 10),
even though their hazy notion of punctuation in English made it harder for them to understand syntax.
We think this is because the younger children were more open to discovery, and unperturbed by making mistakes.
In her PhD thesis, analyzing  four early approaches to integrating computers into schools, \cite{solomon1988computer}, Solomon provides a possible explanation.
She describes the evolution of the four approaches as growing organically from 
differing assumptions about educational psychology.
The most revolutionary, the Logo school, draws on Piaget (as interpreted by Papert) who tells them that children are natural scientists, discovering the rules of their world by bumping up against it,
and later by observing and interacting with older children and adults.
But at some point, organic experience of mathematics fails to motivate the learning of 
arithmetic, and learning objectives must be imposed, which unfortunately suppresses the instinct to make numbers their own.
Papert wanted to extend this intrinsic phase of learning by providing an environment for mathematical
exploration through programming.
Unfortunately, the range of Papert's vision has not been taken up by the Computational Thinking revivalists, because they assume that the way most software is written today must be the best way, and therefore a simplified version thereof will be the best way of teaching children.

\smallskip
Our approach is different.
We have identified achievement in math and especially preparation for high-school algebra as our target,
and have found that we can co-design tools and instruction to meet our initial goals.
To make our curriculum accessible to even younger children,
we have started developing a new iPad app.
We have a lot of anecdotal evidence to support our approach, and advice for improving both tools and instructional design.
Our next steps include acting on that advice, 
fulling implementing touch editing,
enhancing distance mentoring, so that we can scale our program up beyond the schools to which we can (physically) send mentors,
and eventually designing experiments to measure the effect of our instruction on learning.
To main difficulty in designing effective experiments will be the anticipation of
 confounding factors,
such as the mutual reinforcement between programming achievement and spatial awareness 
 \cite{Cooper:2015:SST:2787622.2787728}.

\smallskip
So, if a half century on we are to take up the banner and advocate for the child as scientist,
why should we be more successful?
\begin{description}\setlength{\itemsep}{-1pt}
  \item[Better Hardware]  Whereas Logo started in all caps, because that was what contemporary terminals supported, we have touch interfaces and more processing power in the palm of our hand than a 1970s' supercomputer.  Our iPad app will make better use of this power to construct a more forgiving interface than a text editor. 
  \item[Better Languages]  Logo was designed when structured programming was just beginning to be understood.  We know a lot more now, especially about types.  Having separate types for shapes, stencils, line styles, colours, etc., leads to understandable type errors.  Being able to encode alternatives as alternative constructors makes it easier to map a state diagram into code and back, giving children multiple ways to think about the concept.
  \item[Measurement]  Whereas Logo took children out of school mathematics into Turtle geometry, this made measurement harder, but our approach is based on Cartesian geometry, so it will be easier to measure transfer to the geometry in the curriculum.  Similarly, the semantics of imperative programming and particularly mutable variables conflicts with the usual encoding of word problems into algebra.  The Bootstrap project has found evidence of transfer to general mathematical knowledge \cite{wright2013influence} and word-problem skills \cite{Schanzer:2015:TSS:2676723.2677238}, and we would like to show that we can also make a positive impact in the younger age groups we are teaching.
\end{description}

\smallskip
\emph{We thank the Dean of Engineering, NSERC PromoScience, and Google igniteCS for funding, our many teacher partners for their advice and support over the years, Gordon Goodsman for his contributions to GraphicSVG, the referees for their thoughtful comments and practical suggestions for improving this paper, and the many little scientists for their boundless enthusiasm.}

\bibliographystyle{eptcs}

\begin{thebibliography}{10}
\providecommand{\bibitemdeclare}[2]{}
\providecommand{\surnamestart}{}
\providecommand{\surnameend}{}
\providecommand{\urlprefix}{Available at }
\providecommand{\url}[1]{\texttt{#1}}
\providecommand{\href}[2]{\texttt{#2}}
\providecommand{\urlalt}[2]{\href{#1}{#2}}
\providecommand{\doi}[1]{doi:\urlalt{http://dx.doi.org/#1}{#1}}
\providecommand{\bibinfo}[2]{#2}

\bibitemdeclare{inproceedings}{Achten:2014:TOP:2746325.2746329}
\bibitem{Achten:2014:TOP:2746325.2746329}
\bibinfo{author}{Peter \surnamestart Achten\surnameend},
  \bibinfo{author}{Jurri\"{e}n \surnamestart Stutterheim\surnameend},
  \bibinfo{author}{L\'{a}szl\'{o} \surnamestart Domoszlai\surnameend} \&
  \bibinfo{author}{Rinus \surnamestart Plasmeijer\surnameend}
  (\bibinfo{year}{2014}): \emph{\bibinfo{title}{Task Oriented Programming with
  Purely Compositional Interactive Scalable Vector Graphics}}.
\newblock In: {\sl \bibinfo{booktitle}{Proceedings of the 26Nd 2014
  International Symposium on Implementation and Application of Functional
  Languages}}, \bibinfo{series}{IFL '14}, \bibinfo{publisher}{ACM},
  \bibinfo{address}{New York, NY, USA}, pp. \bibinfo{pages}{7:1--7:13},
  \doi{10.1145/2746325.2746329}.

\bibitemdeclare{mastersthesis}{Helen-Brown-thesis}
\bibitem{Helen-Brown-thesis}
\bibinfo{author}{Helen \surnamestart Brown\surnameend} (\bibinfo{year}{2016}):
  \emph{\bibinfo{title}{MacVenture: An iPad Application Design for Social
  Constructivist E-Learning}}.
\newblock Master's thesis, \bibinfo{school}{McMaster University}.

\bibitemdeclare{article}{TRTR:TRTR68}
\bibitem{TRTR:TRTR68}
\bibinfo{author}{Kathleen~F. \surnamestart Clark\surnameend} \&
  \bibinfo{author}{Michael~F. \surnamestart Graves\surnameend}
  (\bibinfo{year}{2005}): \emph{\bibinfo{title}{Scaffolding Students'
  Comprehension of Text}}.
\newblock {\sl \bibinfo{journal}{The Reading Teacher}}
  \bibinfo{volume}{58}(\bibinfo{number}{6}), pp. \bibinfo{pages}{570--580},
  \doi{10.1598/RT.58.6.6}.

\bibitemdeclare{inproceedings}{Cooper:2015:SST:2787622.2787728}
\bibitem{Cooper:2015:SST:2787622.2787728}
\bibinfo{author}{Stephen \surnamestart Cooper\surnameend},
  \bibinfo{author}{Karen \surnamestart Wang\surnameend}, \bibinfo{author}{Maya
  \surnamestart Israni\surnameend} \& \bibinfo{author}{Sheryl \surnamestart
  Sorby\surnameend} (\bibinfo{year}{2015}): \emph{\bibinfo{title}{Spatial
  Skills Training in Introductory Computing}}.
\newblock In: {\sl \bibinfo{booktitle}{Proceedings of the Eleventh Annual
  International Conference on International Computing Education Research}},
  \bibinfo{series}{ICER '15}, \bibinfo{publisher}{ACM}, \bibinfo{address}{New
  York, NY, USA}, pp. \bibinfo{pages}{13--20}, \doi{10.1145/2787622.2787728}.

\bibitemdeclare{article}{czaplicki2012elm}
\bibitem{czaplicki2012elm}
\bibinfo{author}{Evan \surnamestart Czaplicki\surnameend}
  (\bibinfo{year}{2012}): \emph{\bibinfo{title}{Elm: Concurrent FRP for
  Functional GUIs}}.
\newblock {\sl \bibinfo{journal}{Senior thesis, Harvard University}}.

\bibitemdeclare{techreport}{farewell-to-frp}
\bibitem{farewell-to-frp}
\bibinfo{author}{Evan \surnamestart Czaplicki\surnameend}
  (\bibinfo{year}{2016}): \emph{\bibinfo{title}{A Farewell to FRP}}.
\newblock \bibinfo{type}{Blog post on elm-lang.org}.

\bibitemdeclare{misc}{aljabr2017oed}
\bibitem{aljabr2017oed}
\bibinfo{author}{Oxford~English \surnamestart Dictionary\surnameend}
  (\bibinfo{year}{2017}): \emph{\bibinfo{title}{OED online}}.

\bibitemdeclare{techreport}{UKCSCurriculum2013}
\bibitem{UKCSCurriculum2013}
\bibinfo{author}{Department \surnamestart of~Education\surnameend}
  (\bibinfo{year}{2013}): \emph{\bibinfo{title}{National curriculum in England:
  computing programmes of study}}.
\newblock \bibinfo{type}{Technical Report}, \bibinfo{institution}{UK Department
  of Education}.

\bibitemdeclare{article}{Felleisen:2009:VWC:1538788.1538803}
\bibitem{Felleisen:2009:VWC:1538788.1538803}
\bibinfo{author}{Matthias \surnamestart Felleisen\surnameend} \&
  \bibinfo{author}{Shriram \surnamestart Krishnamurthi\surnameend}
  (\bibinfo{year}{2009}): \emph{\bibinfo{title}{Viewpoint: Why Computer Science
  Doesn't Matter}}.
\newblock {\sl \bibinfo{journal}{Commun. ACM}}
  \bibinfo{volume}{52}(\bibinfo{number}{7}), pp. \bibinfo{pages}{37--40},
  \doi{10.1145/1538788.1538803}.

\bibitemdeclare{inproceedings}{Fisler:2014:RRP:2632320.2632346}
\bibitem{Fisler:2014:RRP:2632320.2632346}
\bibinfo{author}{Kathi \surnamestart Fisler\surnameend} (\bibinfo{year}{2014}):
  \emph{\bibinfo{title}{The Recurring Rainfall Problem}}.
\newblock In: {\sl \bibinfo{booktitle}{Proceedings of the Tenth Annual
  Conference on International Computing Education Research}},
  \bibinfo{series}{ICER '14}, \bibinfo{publisher}{ACM}, \bibinfo{address}{New
  York, NY, USA}, pp. \bibinfo{pages}{35--42}, \doi{10.1145/2632320.2632346}.

\bibitemdeclare{article}{Frey2017254}
\bibitem{Frey2017254}
\bibinfo{author}{Carl~Benedikt \surnamestart Frey\surnameend} \&
  \bibinfo{author}{Michael~A. \surnamestart Osborne\surnameend}
  (\bibinfo{year}{2017}): \emph{\bibinfo{title}{The future of employment: How
  susceptible are jobs to computerisation?}}
\newblock {\sl \bibinfo{journal}{Technological Forecasting and Social Change}}
  \bibinfo{volume}{114}, pp. \bibinfo{pages}{254 -- 280}, \doi{10.1016/j.techfore.2016.08.019}.

\bibitemdeclare{article}{doi:10.2200/S00684ED1V01Y201511HCI033}
\bibitem{doi:10.2200/S00684ED1V01Y201511HCI033}
\bibinfo{author}{Mark \surnamestart Guzdial\surnameend} (\bibinfo{year}{2015}):
  \emph{\bibinfo{title}{Learner-Centered Design of Computing Education:
  Research on Computing for Everyone}}.
\newblock {\sl \bibinfo{journal}{Synthesis Lectures on Human-Centered
  Informatics}} \bibinfo{volume}{8}(\bibinfo{number}{6}), pp.
  \bibinfo{pages}{1--165}, \doi{10.2200/S00684ED1V01Y201511HCI033}.

\bibitemdeclare{article}{doi:10.1093/comjnl/32.2.98}
\bibitem{doi:10.1093/comjnl/32.2.98}
\bibinfo{author}{J.~\surnamestart Hughes\surnameend} (\bibinfo{year}{1989}):
  \emph{\bibinfo{title}{Why Functional Programming Matters}}.
\newblock {\sl \bibinfo{journal}{The Computer Journal}}
  \bibinfo{volume}{32}(\bibinfo{number}{2}), p.~\bibinfo{pages}{98},
  \doi{10.1093/comjnl/32.2.98}.

\bibitemdeclare{article}{kieran2004algebraic}
\bibitem{kieran2004algebraic}
\bibinfo{author}{Carolyn \surnamestart Kieran\surnameend}
  (\bibinfo{year}{2004}): \emph{\bibinfo{title}{Algebraic thinking in the early
  grades: What is it?}}.
\newblock {\sl \bibinfo{journal}{The Mathematics Educator}}
  \bibinfo{volume}{8}(\bibinfo{number}{1}), pp. \bibinfo{pages}{139--151}.

\bibitemdeclare{article}{doi:10.3102/00028312040002353}
\bibitem{doi:10.3102/00028312040002353}
\bibinfo{author}{Valerie~E. \surnamestart Lee\surnameend} \&
  \bibinfo{author}{David~T. \surnamestart Burkam\surnameend}
  (\bibinfo{year}{2003}): \emph{\bibinfo{title}{Dropping Out of High School:
  The Role of School Organization and Structure}}.
\newblock {\sl \bibinfo{journal}{American Educational Research Journal}}
  \bibinfo{volume}{40}(\bibinfo{number}{2}), pp. \bibinfo{pages}{353--393},
  \doi{10.3102/00028312040002353}.

\bibitemdeclare{misc}{gloss}
\bibitem{gloss}
\bibinfo{author}{Ben \surnamestart Lippmeier\surnameend}
  (\bibinfo{year}{2017}): \emph{\bibinfo{title}{The gloss package}}.
\newblock \bibinfo{howpublished}{hackage}.

\bibitemdeclare{inproceedings}{omar2017hazelnut}
\bibitem{omar2017hazelnut}
\bibinfo{author}{Cyrus \surnamestart Omar\surnameend}, \bibinfo{author}{Ian
  \surnamestart Voysey\surnameend}, \bibinfo{author}{Michael \surnamestart
  Hilton\surnameend}, \bibinfo{author}{Jonathan \surnamestart
  Aldrich\surnameend} \& \bibinfo{author}{Matthew~A \surnamestart
  Hammer\surnameend} (\bibinfo{year}{2017}): \emph{\bibinfo{title}{Hazelnut: a
  bidirectionally typed structure editor calculus}}.
\newblock In: {\sl \bibinfo{booktitle}{Proceedings of the 44th ACM SIGPLAN
  Symposium on Principles of Programming Languages}},
  \bibinfo{organization}{ACM}, pp. \bibinfo{pages}{86--99}, \doi{10.1145/3093333.3009900}.

\bibitemdeclare{book}{Papert:1980:MCC:1095592}
\bibitem{Papert:1980:MCC:1095592}
\bibinfo{author}{Seymour \surnamestart Papert\surnameend}
  (\bibinfo{year}{1980}): \emph{\bibinfo{title}{Mindstorms: Children,
  Computers, and Powerful Ideas}}.
\newblock \bibinfo{publisher}{Basic Books, Inc.}, \bibinfo{address}{New York,
  NY, USA}.

\bibitemdeclare{inproceedings}{Schanzer:2015:TSS:2676723.2677238}
\bibitem{Schanzer:2015:TSS:2676723.2677238}
\bibinfo{author}{Emmanuel \surnamestart Schanzer\surnameend},
  \bibinfo{author}{Kathi \surnamestart Fisler\surnameend},
  \bibinfo{author}{Shriram \surnamestart Krishnamurthi\surnameend} \&
  \bibinfo{author}{Matthias \surnamestart Felleisen\surnameend}
  (\bibinfo{year}{2015}): \emph{\bibinfo{title}{Transferring Skills at Solving
  Word Problems from Computing to Algebra Through Bootstrap}}.
\newblock In: {\sl \bibinfo{booktitle}{Proceedings of the 46th ACM Technical
  Symposium on Computer Science Education}}, \bibinfo{series}{SIGCSE '15},
  \bibinfo{publisher}{ACM}, \bibinfo{address}{New York, NY, USA}, pp.
  \bibinfo{pages}{616--621}, \doi{10.1145/2676723.2677238}.

\bibitemdeclare{techreport}{BootstrapReactiveUnit2}
\bibitem{BootstrapReactiveUnit2}
\bibinfo{author}{Emmanuel \surnamestart Schanzer\surnameend},
  \bibinfo{author}{Emma \surnamestart Youndtsmith\surnameend},
  \bibinfo{author}{Kathi \surnamestart Fisler\surnameend},
  \bibinfo{author}{Shriram \surnamestart Krishnamurthi\surnameend},
  \bibinfo{author}{Joe \surnamestart Politz\surnameend} \& \bibinfo{author}{Ben
  \surnamestart Lerner\surnameend} (\bibinfo{year}{2012}):
  \emph{\bibinfo{title}{Bootstrap:Reactive}}.
\newblock \bibinfo{type}{Technical Report}, \bibinfo{institution}{Bootstrap}.

\bibitemdeclare{inproceedings}{Schofield:2014:MCM:2538862.2538901}
\bibitem{Schofield:2014:MCM:2538862.2538901}
\bibinfo{author}{Elizabeth \surnamestart Schofield\surnameend},
  \bibinfo{author}{Michael \surnamestart Erlinger\surnameend} \&
  \bibinfo{author}{Zachary \surnamestart Dodds\surnameend}
  (\bibinfo{year}{2014}): \emph{\bibinfo{title}{MyCS: CS for Middle-years
  Students and Their Teachers}}.
\newblock In: {\sl \bibinfo{booktitle}{Proceedings of the 45th ACM Technical
  Symposium on Computer Science Education}}, \bibinfo{series}{SIGCSE '14},
  \bibinfo{publisher}{ACM}, \bibinfo{address}{New York, NY, USA}, pp.
  \bibinfo{pages}{337--342}, \doi{10.1145/2538862.2538901}.

\bibitemdeclare{article}{silver2008factors}
\bibitem{silver2008factors}
\bibinfo{author}{David \surnamestart Silver\surnameend},
  \bibinfo{author}{Marisa \surnamestart Saunders\surnameend} \&
  \bibinfo{author}{Estela \surnamestart Zarate\surnameend}
  (\bibinfo{year}{2008}): \emph{\bibinfo{title}{What factors predict high
  school graduation in the Los Angeles Unified School District}}.
\newblock {\sl \bibinfo{journal}{Policy Brief}} \bibinfo{volume}{14}.

\bibitemdeclare{book}{solomon1988computer}
\bibitem{solomon1988computer}
\bibinfo{author}{Cynthia \surnamestart Solomon\surnameend}
  (\bibinfo{year}{1988}): \emph{\bibinfo{title}{Computer environments for
  children: A reflection on theories of learning and education}}.
\newblock \bibinfo{publisher}{MIT press}.

\bibitemdeclare{article}{Teitelbaum:1981:CPS:358746.358755}
\bibitem{Teitelbaum:1981:CPS:358746.358755}
\bibinfo{author}{Tim \surnamestart Teitelbaum\surnameend} \&
  \bibinfo{author}{Thomas \surnamestart Reps\surnameend}
  (\bibinfo{year}{1981}): \emph{\bibinfo{title}{The Cornell Program
  Synthesizer: A Syntax-directed Programming Environment}}.
\newblock {\sl \bibinfo{journal}{Commun. ACM}}
  \bibinfo{volume}{24}(\bibinfo{number}{9}), pp. \bibinfo{pages}{563--573},
  \doi{10.1145/358746.358755}.

\bibitemdeclare{article}{vee2013understanding}
\bibitem{vee2013understanding}
\bibinfo{author}{Annette \surnamestart Vee\surnameend} (\bibinfo{year}{2013}):
  \emph{\bibinfo{title}{Understanding Computer Programming as a Literacy}}.
\newblock {\sl \bibinfo{journal}{Literacy in Composition Studies}}
  \bibinfo{volume}{1}(\bibinfo{number}{2}),\doi{10.21623/1.1.2.4}.

\bibitemdeclare{inproceedings}{voelter2014towards}
\bibitem{voelter2014towards}
\bibinfo{author}{Markus \surnamestart Voelter\surnameend},
  \bibinfo{author}{Janet \surnamestart Siegmund\surnameend},
  \bibinfo{author}{Thorsten \surnamestart Berger\surnameend} \&
  \bibinfo{author}{Bernd \surnamestart Kolb\surnameend} (\bibinfo{year}{2014}):
  \emph{\bibinfo{title}{Towards user-friendly projectional editors}}.
\newblock In: {\sl \bibinfo{booktitle}{International Conference on Software
  Language Engineering}}, \bibinfo{organization}{Springer}, pp.
  \bibinfo{pages}{41--61}, \doi{10.1007/978-3-319-11245-9_3}.

\bibitemdeclare{inproceedings}{DBLP:journals/corr/Walck16}
\bibitem{DBLP:journals/corr/Walck16}
\bibinfo{author}{Scott~N. \surnamestart Walck\surnameend}
  (\bibinfo{year}{2016}): \emph{\bibinfo{title}{Learn Quantum Mechanics with
  Haskell}}.
\newblock In: {\sl \bibinfo{booktitle}{Proceedings of the 4th and 5th
  International Workshop on Trends in Functional Programming in Education,
  {TFPIE} 2016, Sophia-Antipolis, France and University of Maryland College
  Park, USA, 2nd June 2015 and 7th June 2016.}}, pp. \bibinfo{pages}{31--46}, \doi{10.4204/EPTCS.230.3}.

\bibitemdeclare{article}{Wing:2006:CT:1118178.1118215}
\bibitem{Wing:2006:CT:1118178.1118215}
\bibinfo{author}{Jeannette~M. \surnamestart Wing\surnameend}
  (\bibinfo{year}{2006}): \emph{\bibinfo{title}{Computational Thinking}}.
\newblock {\sl \bibinfo{journal}{Commun. ACM}}
  \bibinfo{volume}{49}(\bibinfo{number}{3}), pp. \bibinfo{pages}{33--35},
  \doi{10.1145/1118178.1118215}.

\bibitemdeclare{inproceedings}{wright2013influence}
\bibitem{wright2013influence}
\bibinfo{author}{Geoff \surnamestart Wright\surnameend}, \bibinfo{author}{Peter
  \surnamestart Rich\surnameend} \& \bibinfo{author}{Robert \surnamestart
  Lee\surnameend} (\bibinfo{year}{2013}): \emph{\bibinfo{title}{The influence
  of teaching programming on learning mathematics}}.
\newblock In: {\sl \bibinfo{booktitle}{Society for Information Technology \&
  Teacher Education International Conference}},
  \bibinfo{organization}{Association for the Advancement of Computing in
  Education (AACE)}, pp. \bibinfo{pages}{4612--4615}.

\bibitemdeclare{inproceedings}{Yorgey:2012:MTV:2364506.2364520}
\bibitem{Yorgey:2012:MTV:2364506.2364520}
\bibinfo{author}{Brent~A. \surnamestart Yorgey\surnameend}
  (\bibinfo{year}{2012}): \emph{\bibinfo{title}{Monoids: Theme and Variations
  (Functional Pearl)}}.
\newblock In: {\sl \bibinfo{booktitle}{Proceedings of the 2012 Haskell
  Symposium}}, \bibinfo{series}{Haskell '12}, \bibinfo{publisher}{ACM},
  \bibinfo{address}{New York, NY, USA}, pp. \bibinfo{pages}{105--116},
  \doi{10.1145/2364506.2364520}.

\end{thebibliography}

\end{document}